# COUSTIC: Combinatorial Double Auction for Crowd sensing Task Assignment in Device-to-Device Clouds

Yutong Zhai[1], Liusheng Huang[1], Long Chen[1], Ning Xiao[1], Yangyang Geng[1]

[1] School of Computer Science and Technology, University of Science and Technology of China, China
zyt1996@mail.ustc.edu.cn, lshuang@ustc.edu.cn,
lonchen@mail.ustc.edu.cn, xiaoning@mail.ustc.edu.cn,
geng325@mail.ustc.edu.cn

**Abstract.** With the emerging technologies of Internet of Things (IOTs), the capabilities of mobile devices have increased tremendously. However, in the big data era, to complete tasks on one device is still challenging. As an emerging technology, crowdsourcing utilizing crowds of devices to facilitate large scale sensing tasks has gaining more and more research attention. Most of existing works either assume devices are willing to cooperate utilizing centralized mechanisms or design incentive algorithms using double auctions. Which is not practical to deal with the case when there is a lack of centralized controller for the former, and not suitable to the case when the seller device is also resource constrained for the later. In this paper, we propose a truthful incentive mechanism with combinatorial double auction for crowd sensing task assignment in device-to-device (D2D) clouds, where a single mobile device with intensive sensing task can hire a group of idle neighboring devices. With this new mechanism, time critical sensing tasks can be handled in time with a distributed nature. We prove that the proposed mechanism is truthful, individual rational, budget balance and computational efficient. Our simulation results demonstrate that combinatorial double auction mechanism gets a 26.3% and 15.8% gains in comparison to existing double auction scheme and the centralized maximum matching based algorithm respectively.

**Keywords:** Mobile Crowd Sensing, Device-to-Device Clouds, Combinatorial Double Auction, Task Allocation.

# 1  Introduction

Throughout the last a few years, mobile devices like smart phones, laptops and ipads have become proliferation in people's daily life. They can generate massive information about the environment by themselves for sensing the physical world. Mobile crowd sensing [18], which defined as individuals with sensing and computing devices collectively share data and extract information to measure and map phenomena of common interest, is more and more popular in the evolution of the Internet of Things (IoT).

In general, mobile crowd sensing classifies as personal sensing and community sensing. Personal sensing always senses one simple task, like the monitoring of movement patterns of an individual for personal record-keeping or health care reasons. Community sensing defines as a cognitive sensing mode that requires many individual devices participated together which is suitable for large-scale phenomenon, such as intelligent transportation systems.

Although new devices are becoming more and more powerful, users take for granted the ability of their resources to perform complicated sensing tasks like air quality monitoring [19], traffic information mapping [20] and public information sharing [21]. However, the complicated sensing tasks are constrained by the limited computational, energy and data resources. In order to ensure the normal operation of sensing applications, we can consider a set of mobile devices forming what we called Device-to-Device (D2D) Clouds [1]. Device-to-device cloud are consisted of a set of mobile devices and a cluster head or an access point (AP), shown in Figure 1.

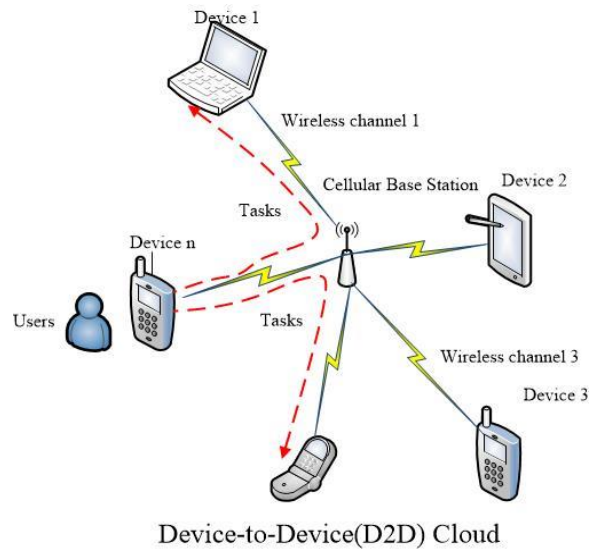

Fig. 1. One example of device-to-device (D2D) cloud

Users or mobile devices may generate many large scale sensing tasks which is difficult to process in a short time efficiently by a single device. By the use of crowdsourcing we can offload the task to nearby devices. Therefore, task allocation is a key issue in mobile crowd sensing. There have been two incentive methods for task allocation, one called centralized algorithm and the other was distributed algorithm. In the distributed algorithm, game theory and auction are two common ways used in task allocation. Recently, auction is becoming a popular method for solving task allocation problem in mobile crowd sensing [10]. In general, there are four types auction——single auction, double auction, combinatorial auction and combinatorial double auction. An auction involving both buyers and sellers is called double auction [12], combinatorial auction mechanism is first used for airport time slot allocation [13], combinatorial double auction is first proposed in [14] for a market with public goods. Existing studies for crowd sensing such as [22] [23] [24] are based on double auction mechanisms. In the double auction mechanisms, one task can only be offloaded to one devices, this may cause big delay while processing a large scale tasks. Therefore, it is necessary to use a combinatorial double auction (CDA) mechanism for the task allocation by the use of crowdsourcing.

Designing a CDA mechanism for mobile crowd sensing has two major challenges. For all the participants, both the sensing tasks and the devices, the mechanisms is essential to bring benefits and be fairly to them. Another major challenge is how to allocate the sensing tasks, in the general case, finding the optimal task allocation solution is a NP-hard problem that cannot be solved in polynomial time.

In this paper, we mainly consider a Combinatorial dOuble aUction for taSk assignmenT in device-to-device Crowdsourcing clouds (COUSTIC) problem. To tackle the problem and the two challenges mentioned below, we try to design an efficient combinatorial double auction mechanism for task allocation in D2D clouds. The major contributions of this paper can be summarized as follows:

- We first analyze the situation where there are several tasks and several mobile devices in the D2D clouds, then we formulate the task assignment as a combinatorial double auction problem into an integer programming problem. That we design a polynomial time greedy algorithm to reduce the computational complexity and solve the utility maximization problem.
- The proposed greedy algorithms are efficient because of their polynomial time complexity running time. We prove that the proposed auction algorithms are truthful, budget balanced and they are individual rational.
- We build the D2D clouds model and the combinatorial double auction (CDA) model, and compare our model with the double auction, random allocation, and the maximum matching allocation mechanism. Simulation results show that our model get a good result, it can be treat as a new method for the task assignments in D2D clouds.

The rest of the article is organized as follows. Section 2 introduces the related works on auction schemes for task allocation. Section 3 describes the system model and formulate the problem. Section 4 presents the algorithm for task assignment and pricing payments, then prove the several properties of the auction. Section 5 presents the simulation results. Finally, we conclude the paper in Section 6.

## 2   Related work

In recent years, many works of D2D clouds have been proposed. For example, A .Mitbaa et.al.[1][2] develop a computational offloading scheme that maximize the lifetime of the ensemble of mobile device. However, in D2D clouds, not all the devices are willing to offload extra tasks without payment. K .Habak et.al.[3] consider how a collection of co-located devices can be orchestrated to provide a cloud service at edge and present the FemtoClouds system. X .Wang et.al.[4] constructs the task assignment in D2D clouds based on double auction mechanisms. Auction is a popular trading mechanism that can allocate resources between the buyers and the sellers. In auction, prices of the resources are determined by the task's (buyers) willingness to pay and the devices (sellers) would not face the uncertainly of the best resource price [5].

Double auction model is commonly used in task allocation problems. X.Wang et.al.[4] proposed a double auction model in D2D clouds. They consider both situation of homogeneous and heterogeneous, and design corresponding algorithms to solve the task assignment problem and payment scheme. D .Yang et.al.[12] also announced a truthful double auction scheme for cooperative communications (TASC) in D2D clouds, the TASC model is also individual rationality, budget balanced and truthfulness. However, the models mentioned above do not consider the situation of crowdsourcing for a large scale task.

Mobile Crowd sensing is closely connected with auction, several incentive auction model were proposed to attract more user participated in providing service [15] [16] [17]. Similar to the crowdsourcing service mechanism, Combinational double auction mechanism is also an incentive mechanism which offload the task to other devices for reducing latency and improving data quality. It is also another feasible way to solve problems in different scenes. L .Chen et.al.[6] proposed a novelty CDA mechanism for spectrum allocation in cognitive radio network, G .Baranwal et.al.[7] also announced a fair CDA model for resource allocation in cloud computing. K .Xu proposed that CDA is also adopted for mobile cloud computing markets [8]. P .Samimi .et.al[9] also announced a CDA model in cloud computing for resource allocation, but the model does not satisfies some economic properties, like individual rationality. However, the model proposed in [9] did not achieve individual rationality, and the model proposed in [8] is too complicated to be suitable for a simple D2D cloud situation.

## 3   Problem Definition

In this section, we first introduce the model of D2D clouds and the model of combinatorial double auction (CDA). Then we formulate the problem and present the four economic properties of the auction.

### 3.1 Device-to-Device Cloud Model

We consider a situation in a Device-to-Device (D2D) Cloud, combined with an access point (AP) and a set of mobile users. In a D2D cloud, we suppose that $D = \{d_1, d_2, \ldots d_n\}$ are the mobile users willing to participate in task assignments. Each user in D may carry the task by itself, or it may be idle, or it can participate in the task assignment.

In general, we believe that m tasks $T = \{T_1, T_2, \ldots T_m\}$ are created during the use of mobile devices, and we assume that each task are consisted of multiple subtasks. For each sub-task $T_i, i \in \{1,2, \ldots m\}$ is denoted by $T_i = \{\theta_{i,1}t_1, \theta_{i,2}t_2, \ldots \theta_{i,k}t_k\}$, $\theta_{i,j}$ denotes the demand for each type of tasks. And $v_i$ is the true valuation for the computing process of the sub-task $T_i$. The problem we need to solve is the allocation of these applications.

For a group of mobile device in a D2D cloud network, due to the heterogeneity of mobile devices, different mobile devices have different computing resources, storage capacity, and so on. Therefore, the types and quantities of tasks that can be run on each device are also different. We assume as follows: each devices j has its own free resources $R_j$, the total free resources of n devices are represented as: $\vec{R} = \{R_1, R_2, \ldots R_n\}$.

### 3.2 Combinatorial Double Auction Model

In this system, there are several sensing tasks (buyers) which need to be processed efficiently and devices (sellers) which have a large amount of computing resources, and the access point serves as an auctioneer. Each task submits a bid to the access point, denoted as $B_i = \{\vec{\theta_i}, v_i\}$ where $\vec{\theta_i} = \{\theta_{i,1}, \theta_{i,2}, \ldots \theta_{i,k}\}$ represents the demand amount of every subtask. $v_i$ denotes the truthful value evaluated by the ith task.

Consider about all mobile devices (sellers), each device submits a bid $S_j$ to the auctioneer, $j \in \{1,2, \ldots n\}$. $S_j = \{\vec{s_j}, w_j, c_j\}$ where $\vec{s_j} = \{s_{j,1}, s_{j,2}, \ldots s_{j,k}\}$. $s_{j,i}$ represents the ith type resource owned by device j, $\vec{s_j}$ denotes a combination of resource provided by device j. $w_j$ represents the maximum number that can provide this type of service. $c_j$ denotes the truthful value evaluated by the ith device's cost.

While the access point (AP) get all bids from buyers and sellers. It acts as an auctioneer to determine who has failed or wined in the auction, and expresses a task assignment scheme at the same time. The scheme includes one task offloaded to which devices and each devices offload how many. At the same time, the auctioneer computes the charge and payment by a pricing model. Then get charge from tasks and pay corresponding money to the sellers. The flow of the proposed CDA model shows in Fig.2.

### 3.3 Problem Formulation

In this work, one application is consisted of multiple tasks capable to be divided and to be distributed to multiple mobile devices for parallel computing in the auction model. And the extra cost of distributed computing denoted as $e_i = \alpha * t_i + \beta$ where α and β are positive constant.

Let M denote the set of tasks in the D2D cloud and |M|=m. Let N denote the set of mobile users in the D2D cloud and |N|=n. Obviously, the access point (AP) formed an allocation matrix. The matrix is a M*N matrix and is denoted as $x_{mn}$.

Each element in matrix X can be denoted as

$$x_{ij} = \begin{cases} 1, & \text{if device } j \text{ offload part of the ith task} \\ 0, & \text{otherwise} \end{cases} \quad (1)$$

Since the task is offloaded to each device, the sum of the requested resources cannot exceed the maximum resources of the device. Thus

$$\sum_{k=1}^{m} m_{kj} \leq w_j, j \in \{1, 2, \ldots n\} \quad (2)$$

At the same time, each task gets enough resources from several devices. Hence

$$\sum_{k=1}^{n} x_{ik} * \vec{s_k} \geq \vec{\theta_i}, i \in \{1, 2 \ldots m\} \quad (3)$$

Let $trade\ price_i$ denotes the final payment for the ith task. And $trade\ price_j$ denotes the final payment for the jth device. Then we can get the utility for both the buyers and the sellers. The utility of ith task (buyer) is:

$$utility_i^b = \begin{cases} (v_i - e_i) - trade\ price_i, & \text{if ith task win the auction} \\ 0, & \text{if ith task lose the auction} \end{cases} \quad (4)$$

The utility of jth device (seller) is:

$$utility_j^s = \begin{cases} trade\ price_j - \sum_{k=1}^{m} x_{kj}, & \text{if jth device win the auction} \\ 0, & \text{if jth device lose the auction} \end{cases} \quad (5)$$

Therefore, given the bids from the tasks and the devices, firstly the access point (AP) needs to determine the allocation matrix X, by the allocation matrix, we can also get the payment and the charge for the buyers and sellers. The problem to determine X can be formulated as follows:

$$\max(\sum_{i=1}^{m} utility_i^b + \sum_{j=1}^{n} utility_j^s) \quad (6)$$

subject to:

$$\sum_{i=1}^{m} x_{ij} \leq w_j, \forall d_j \in D \quad (7)$$

$$\sum_{j=1}^{n} x_{ij} * \vec{s_j} \geq \vec{\theta_i}, \forall t_i \in T \quad (8)$$

$$x_{ij} \in \{0,1\}, i \in \{1, 2 \ldots m\} \text{ and } j \in \{1, 2 \ldots n\} \quad (9)$$

Refer to the work in [11], the above problem of determine allocation matrix is an NP-hard problem. The optimal solution is unable to be obtained in polynomial time.

The problem has both capacity constraint and deadline limitation. Unlike traditional double auction model [4][12], the use of multiple combinations increases the complexity and difficulty.

### 3.4 Economic Properties

Our goal is to design a combinatorial double auction mechanism, and the mechanism is able to achieve the three economic properties in polynomial time: individual rationality, budget balance and truthfulness.
**Individual Rationality.** A combinatorial auction is individually rational if no winner's utility is negative.
**Budget Balance.** The auctioneer's revenue is not negative. The property is utilized to motivate the auctioneer to participate in the auction.
**Truthfulness.** The bid submitted by each buyer and seller should be truthful. No one can get more utility by submitted a fake bid.
**Computational efficiency.** The mechanism or algorithm we propose should be solved in polynomial time.

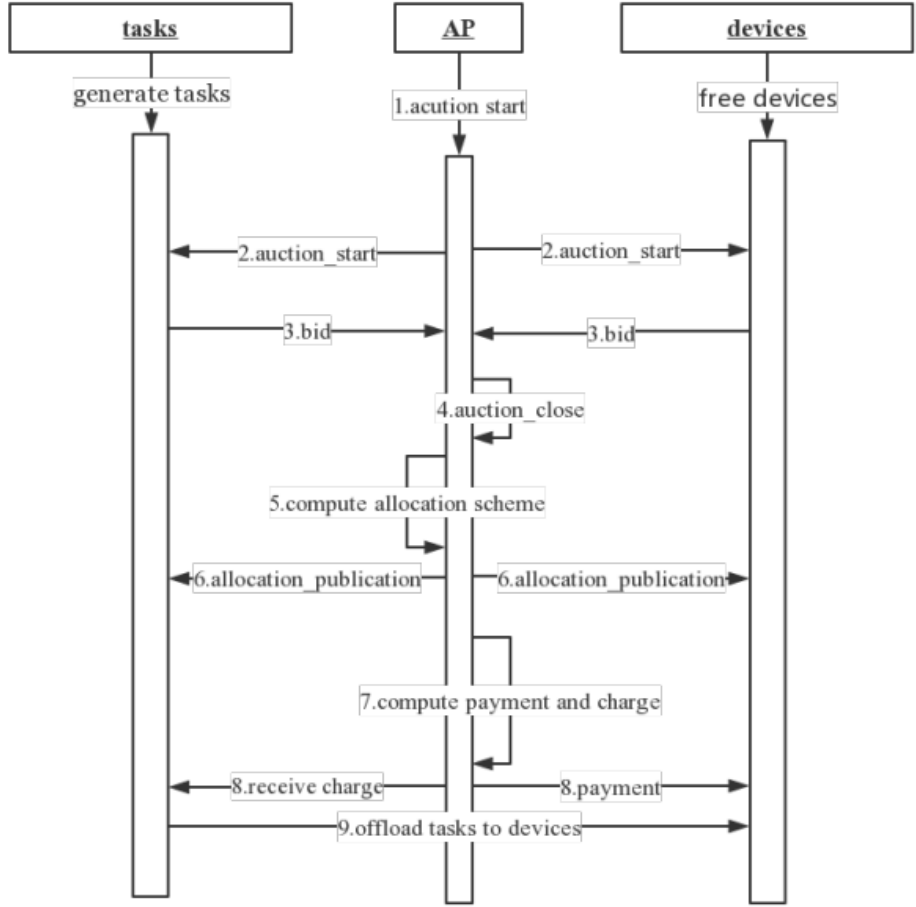

**Fig. 2.** The flow of communications among the combinatorial double auction model in MDCs.

## 4 Greedy Allocation Mechanism

In this section, we present a greedy algorithm to solve the NP-hard problem ([Eq.6] to [Eq.9]). Although the optimal solution of an NP-hard problem is impossible to get in polynomial time, it is feasible to get an approximate optimal solution. The bid density is utilized to push bids for an efficient allocation, and the bid density must be a monotonic function. The bid density should reflect the value of sub-tasks and resources, which is a function about v or c. Different from the bid density used in [9], we define the bid density of tasks and devices as follows:

$$bd\_task_i = \sqrt{\sum_{j=1}^{k} \theta_{i,j}^2 + v_i^2}, \ i \in \{1,2\ldots m\} \qquad (10)$$

$$bd\_device_j = \sqrt{\sum_{j=1}^{k} 1/s_{i,j}^2 + c_j^2}, \ j \in \{1,2 \dots n\} \tag{11}$$

Therefore, we have

$$\sqrt{v_1^2 + \sum_{j=1}^{k} \theta_{1,j}^2} \geq \sqrt{v_2^2 + \sum_{j=1}^{k} \theta_{2,j}^2} \geq \cdots \geq \sqrt{v_m^2 + \sum_{j=1}^{k} \theta_{m,j}^2} \tag{12}$$

$$\sqrt{c_1^2 + \sum_{j=1}^{k} 1/s_{1,j}^2} \geq \sqrt{c_2^2 + \sum_{j=1}^{k} 1/s_{2,j}^2} \geq \cdots \geq \sqrt{c_n^2 + \sum_{j=1}^{k} 1/s_{n,j}^2} \tag{13}$$

Then, we design the allocation model and the pricing model are shown in algorithm 1 and algorithm 2, and the complexity of two models are both O(mn). And prove the pricing and the allocation model satisfy several economic properties.

### 4.1 Allocation Model

The allocation model is a scheme or an algorithm to determine which task should be assigned to which device. In a D2D cloud, bids of tasks and mobile devices are submitted to the access point (AP). According to the definition of bid density [Eq.12, 13], the bids are sorted in ascending order and descending order by the access point (AP). The AP determines the allocation matrix by a greedy strategy which describes in Algorithm 1(see Fig.3).

```
Algorithm 1
```
**Input:** M, N, B, S   //The number of tasks M, the number of devices N, bids of the tasks B
// bids of the devices S
**Output:** X //allocation matrix X
1: Calculate and sort the bids of tasks and devices
2: E = {∅}
3: Normalized v and θ between (0,1)
4: for i = 1 to M do
5:   $bd\_task_i = \sqrt{\sum_{j=1}^{k} \theta_{i,j}^2 + v_i^2}$
6:   E = E ∪ {$bd\_task_i$}
7: end for
8: sort bid densities in E in descending order
9: F = {∅}
10: Normalized s and c between (0,1)
11: for j = 1 to N do
12:   $bd\_device_j = \sqrt{\sum_{j=1}^{k} 1/s_{i,j}^2 + c_j^2}$
13:   F = F ∪ {$bd\_device_j$}
14: end for
15: sort bid densities in F in ascending order
16: initialize allocation matrix X
17: $X_{mn} = 0$

```
18: for e = 1 to |E| do
19:    for f = 1 to |F| do
20:       if w > 0 and v > c and θ > 0: // enough cost and enough resources
21:          X_ef = 1, w = w − 1, θ = θ − s
22:       end if
23:    end for
24:    if θ < 0: //announce the eth task win the auction
25:    end if
26: end for
27: get allocation matrix X
```

Fig. 3. The greedy allocation algorithm

### 4.2   Pricing Model

We then introduce the pricing model which decides the payment of buyers and the charge of sellers. The payment should be enough fairy. Similar to the work of [5] [7], while the allocation matrix is fixed, the winning resources of each buyer is also fixed as follows:

$$resource_i = \sum x_{ij} * \sum s_j \ , \ i \in \{1,2 \ldots m\}, \ j \in \{1,2 \ldots n\} \tag{14}$$

We define the per unit price for resources as follows:

$$per\ price\_task_i = \frac{v_i}{\sum x_{ij} * \sum s_j} \ , i \in \{1,2 \ldots m\}, \ j \in \{1,2 \ldots n\} \tag{15}$$

$$per\ price\_device_j = \frac{c_j}{\sum s_j} \ , i \in \{1,2 \ldots m\}, \ j \in \{1,2 \ldots n\} \tag{16}$$

Next, the average price matrix P was calculated based on the average value of [Eq.15] and [Eq.16], then we use the algorithm 2(see Fig.4) to calculate corresponding payments by the price matrix, then the auctioneer would send the payment and the charge to both buyers and sellers.

```
Algorithm 2
```
**Input:** X, B, S  //The allocation matrix X, bids of the tasks B, bids of the devices S
**Output:** P, $trade\ price\_task_i$, $trade\ price\_device_j$

1: Calculate the per unit pricing matrix
2: for the winner of the buyers
3:     $per\ price\_task_i = \frac{v_i}{\sum x_{ij} * \sum s_j}$
4: end for
5: for the winner of the sellers
6:     $per\ price\_device_j = \frac{c_j}{\sum s_j}$
7: end for
8: initialize pricing matrix P
9: for the winner of the buyers

```
10:    for the winner of the sellers
11:        P_ij = (per price_task_i + per price_device_j)/2
12:        if per price_device_j ≤ P_ij ≤ per price_task_i
13:        end if
14:        else P_ij = per price_device_j
15:    end for
16: end for
17: calculate the payment and the charge
18: for the winner of the buyers
19:     trade price_task_i = Σ_{m=1}^{k} P_im * s⃗_j * x_im
20: end for
21: for the winner of the sellers
22:     trade price_device_j = Σ_{i=1}^{k} P_kj * s⃗_j * x_kj
23: end for
24: get trade price_task_i, trade price_device_j
```

Fig. 4. The greedy allocation algorithm

### 4.3 Algorithm Analysis and Auction Properties

We now analyze the properties of the allocation and pricing model, including computational efficiency, individual rationality, budget balance and truthfulness.

**Time complexity analysis.**

**Theorem 1:** The time complexity of the allocation model is $O(m\log m + n\log n + mn)$ and the pricing model is $O(mn)$.

**Proof 1:** For the allocation model, according to algorithm 1, the complexity for sorting the bid density is $O(m\log m + n\log n)$, and the generate allocation matrix phase is $O(mn)$. Therefore, the complexity of the allocation is $O(m\log m + n\log n + mn)$. There are two loops in algorithm 2, thus the complexity of the pricing model is $O(mn)$ obviously. Both of them can be completed in polynomial time.

**Individual Rationality.**

**Theorem 2:** For each buyer and seller, its utility is not negative.

**Proof 2:** According to [Eq.4], the utility of the winner of the buyer is equal to the valuation plus the trade price (payment), and the utility of the buyer lose the auction equals to zero. In algorithms 1, if the buyer has no extra value to support the cost, it will lose the auction. Therefore, the winner have enough value to support the cost, and the final payment of buyers is higher than the cost but lower than the valuation, so the utility of each buyer is not negative.

According to [Eq.5], for the winner of the seller, because of the per unit pricing is more expensive than the cost, the per unit utility of each seller is not negative, then the utility is also not negative. And the utility of the seller lose the auction equals to zero. Therefore, the utility of each seller is not negative.

**Budget Balance.**

**Theorem 3:** The auctioneer in the auction is budget balanced.

**Proof 3:** When the auctioneer gets all the bids, it will determine the winner and payment of the auction by the allocation model and pricing model, the value of utility of the auctioneer equals to the difference between the payments received from all buyers and the charges payed to all sellers. Therefore we have

$$utility_{auctioneer} = \sum trade\ price\_task_i - \sum trade\ price\_device_j$$

Because of the per cost of buyers is higher than sellers, so the buyer's sum of payments is greater than the seller's sum of payments, and the utility of the auctioneer is not negative, in other words, the auctioneer in the auction is budget balanced.

**Truthfulness.**

**Theorem 4:** The greedy allocation and pricing mechanisms are truthful.

**Proof 4:** While the allocation matrix X is fixed, both the winner of buyer and the seller in the auction get the most efficient allocation. By the pricing strategy, the revenue is fixed and maximized. Buyers (sellers) cannot improve their own utility by submitting a fake bid.

For the lose buyer failed in the auction. If the buyer submits a lower bid, he will still cause his own auction to fail. In this case, the utility is still zero, if a higher bid is submitted, it will cause you to obtain resources, but the resulting utility is negative.

For the lose seller failed in the auction. If a higher bid is submitted, it still fails and the utility is still zero, if the bid is submitted with a lower bid, it will cause him to obtain the buyer, but the actual cost is too large and the utility will be negative.

In summary, the greedy allocation and pricing mechanisms are truthful.

## 5  Simulation Setup and Experimental Results

In this section, we evaluate the performance of our algorithms.

The presented evaluation metrics are (1): the individual rationality of applications and mobile devices; (2): the percentage served users, which is the ratio of the number of winning users to the total number of users and the (average) utility of tasks and devices. We will also compare our mechanism with an existing double auction mechanism and a maximum matching mechanism to show our improvement.

By default we assume there are six sensing applications and five devices in a device-to-device (D2D) cloud, each task can be divided into two different types of subtasks, task1 and task2. We list the task amount and the corresponding bid in Table 1, each mobile device has own unused resources, as shown in Table 2. According to the evaluation function eq. [10] [11], we compute the bid density for tasks (buyers) and devices (sellers). And we assume that α=0.01 and β=0.05.

Table 1. Bid density of the tasks (buyers)

| Tasks | Task1($\theta_1$) | Task2($\theta_2$) | v | Bid density |
|---|---|---|---|---|
| T1 | 30 | 30 | 13 | 1.732051 |
| T2 | 30 | 20 | 12 | 1.563472 |
| T3 | 25 | 25 | 12 | 1.545603 |
| T4 | 30 | 20 | 11 | 1.511530 |
| T5 | 15 | 15 | 10 | 1.092906 |
| T6 | 10 | 15 | 9 | 0.961047 |

Table 2. Bid density of the devices (sellers)

| Devices | Resource1 | Resource2 | W | v | Bid density |
|---|---|---|---|---|---|
| D1 | 3 | 6 | 6 | 1 | 0.885689 |
| D2 | 5 | 5 | 6 | 1.2 | 0.958458 |
| D3 | 8 | 6 | 6 | 1.5 | 1.283333 |
| D4 | 10 | 8 | 5 | 2 | 1.670366 |
| D5 | 9 | 9 | 4 | 2 | 1.676305 |

### 5.1 Performance on Individual Rationality

We first investigate the performance of the proposed mechanisms on individual rationality.

As shown in Fig.4, we see the submitted bid of the tasks (buyers) are more than the final payment and the submitted bid of mobile devices (sellers) are less than the final payment, from Fig.5 and Fig.6, we can see both the sellers and the buyers can get positive utility. In other words, the algorithms we proposed can achieve the individual rationality.

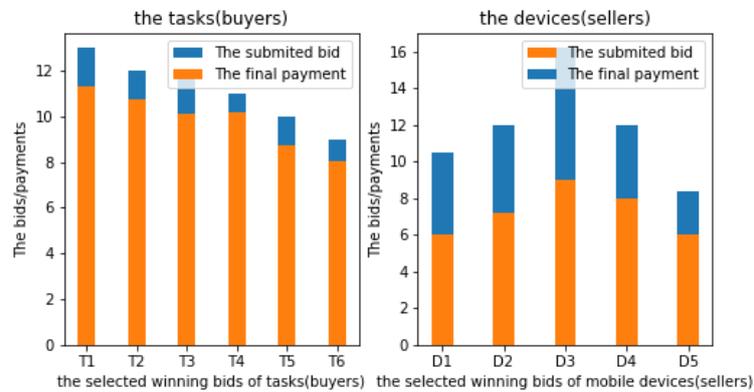

Fig. 3. Performance on Individual Rationality

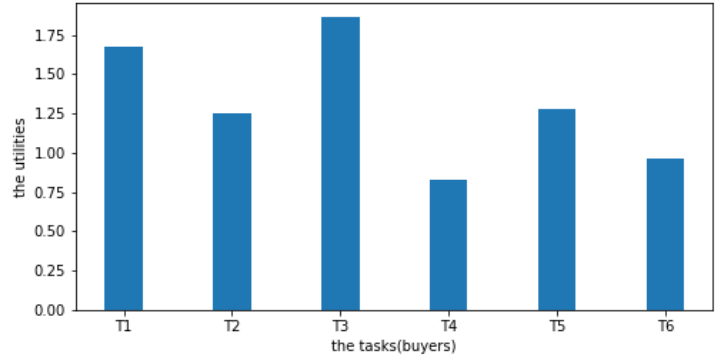

Fig. 4. Utility of Tasks (Buyers)

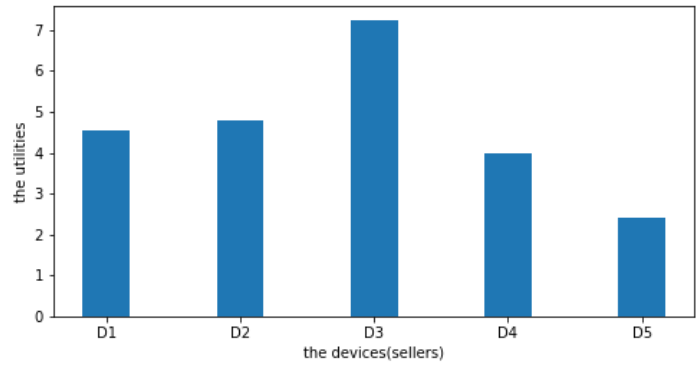

Fig. 5. Utility of Mobile Devices (Sellers)

### 5.2 Percentage Served and Total Utility

We second compare our model with the double auction model in [4], the random allocation scheme, and the maximum matching scheme by the percentage served, total utility and average utility. Results for the three metrics are illustrated in Fig. 7(a), Fig. 7(b) and Fig. 7(c), respectively.

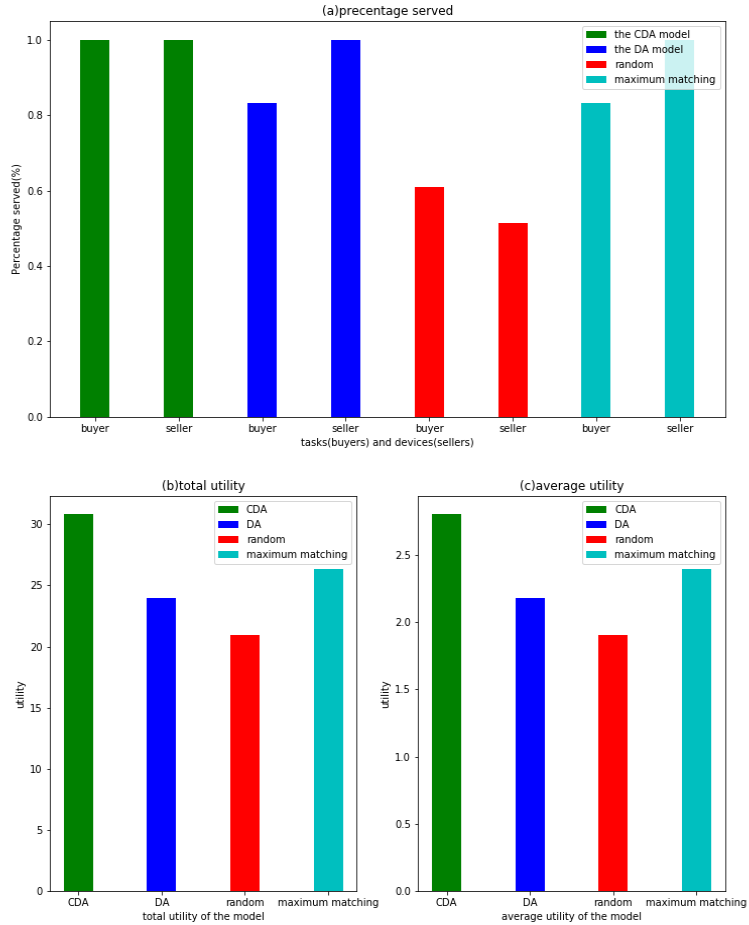

**Fig. 7.** Total Utility and Average Utility

As shown in Fig. 7(a), we observe the combinatorial double auction (CDA) get a better percentage served than the double auction, in the combinatorial double auction (CDA) model, we can see six buyers and five sellers win the auction; but in the double auction model, only five buyers and five sellers win the auction, one buyer lost the auction. We run two thousand times random allocation model, it can be seen the random scheme get an average of 3.686 tasks and 3.074 devices get the task offloading. And the maximum matching schemes also can not get full use of tasks and devices. Then we consider about the utility of the auction. Fig. 7(c) has the same proportional relation and tendency with Fig. 7(b), which means the total utility and the average utility is a proportional relationship, and the ratio is the total number of sellers and buyers. The combinatorial double auction (CDA) model get a 26.3% and 15.8% benefits than the double auction (DA) model and maximum matching mechanism. In the CDA model, a task is divided into multiple subtasks and offloaded to different device to execute, and a device is able to offload multiple tasks synchronously, this accelerates the auction process and improve the efficiency of the devices.

## 6      Conclusion and Future work

In this paper, we discuss sensing task assignment problem in D2D clouds and we propose a combinatorial double auction mechanism to assign the sensing tasks to different mobile devices for distribute processing or parallel processing, where the tasks act as the buyers and the devices as the sellers. Then we analyze the economic properties. The simulation results show that the mechanisms can achieve a fairy performance than the traditional task allocation schemes and also achieve economic properties mentioned earlier. But we have not consider the dependencies between tasks. In future work, we will consider to add the task dependency to parallel tasks and the release of mobile device resources. And it is also able to consider the survival time of mobile devices in the auction mechanisms.

## References

1. A. Mtibaa, A. Fahim, K. A. Harras, and M. Ammar.: "Towards resource sharing in mobile device clouds: Power balancing across mobile devices." Proceedings of the second edition of the MCC workshop on Mobile cloud computing (MCC), pp. 51–56 (2013).
2. A. Mtibaa, K. A. Harras, A. Fahim.: "Towards Computational Offloading in Mobile Device Clouds." In: Proceeding CLOUDCOM '13 Proceedings of the 2013 IEEE International Conference on Cloud Computing Technology and Science Volume 01, pp. 331–338 (2013).
3. K. Habak, M. Ammar, K. A. Harras.: "FemtoClouds: Leveraging Mobile Devices to Provide Cloud Service at the Edge." In: Proceeding CLOUD '15 Proceedings of the 2015 IEEE 8th International Conference on Cloud Computing, pp. 9–16 (2015).
4. X. Wang, X. Chen, W. Wu.: "Towards Truthful Auction Mechanisms for Task Assignment in Mobile Device Clouds." IEEE Conference on Computer Communications(INFOCOM), pp. 1-9 Atlanta, USA, (2017).
5. R. PrestonMcAfee.: "A dominant strategy double auction." Journal of Economic Theory, Volume 56, Issue 2, pp. 434-450 (1992).
6. L. Chen, L. Huang, Z. Sun, H. Xu, H. Guo.: "Spectrum combinatorial double auction for cognitive radio network with ubiquitous network resource providers." IET Communications Volume 9, pp. 2085–2094 (2015).
7. G. Baranwal, D. P. Vidyarthi.: "A fair multi-attribute combinatorial double auction model for resource allocation in cloud computing". Journal of Systems and Software Volume 108, pp. 60–76 (2015).
8. K. Xu, Y. Zhang, X. Shi, H. Wang, Y. Wang.: "Online combinatorial double auction for mobile cloud computing markets." IEEE 33rd International Performance Computing and Communications Conference(IPCCC), pp. 1–8 (2014).
9. P. Samimi, Y. Teimouri, M. Mukhtar.: "A combinatorial double auction resource allocation model in cloud computing." Information Sciences, (2014).

10. Z. Feng, Y. Zhu, Q.Zhang.: "TRAC: Truthful auction for location-aware collaborative sensing in mobile crowdsourcing." IEEE Conference on Computer Communications(INFOCOM), pp. 1231 - 1239 (2014).
11. W. Hu, H. Huang, Y. Sun.: "DATA: A double auction based task assignment mechanism in crowdsourcing systems." 8th International Conference on Communications and Networking in China (CHINACOM), pp. 172–177 (2013).
12. D. Yang, X. Fang, G. Xue.: "Truthful auction for cooperative communications." MobiHoc '11 Proceedings of the Twelfth ACM International Symposium on Mobile Ad Hoc Networking and Computing, (2011).
13. S.J. Rassenti, V.L. Smith, R.L. Bulfin.: "A combinatorial auction mechanism for airport time slot allocation." The Bell Journal of Economics, Volume 13, Issue 2, pp. 402–417 (1982).
14. S. Ba, J. Stallaert, AB. Whinston.: "Optimal investment in knowledge with in a firm using a market-mechanism." Management Science, pp. 1203–1219 (2001).
15. B. Satzger, H. Psaier, D. Schall: "Auction-based crowdsourcing supporting skill management." Information System (2013).
16. D. yang, G. Xue, X. Fang.: "Crowdsourcing to smartphones: incentive mechanism design for mobile phone sensing." Proceedings of the 18th annual international conference on Mobile computing and networking (MOBICOM), pp. 173–184 (2012).
17. H. Shah, VWS. Wong.: "Profit maximization in mobile crowdsourcing: A truthful auction mechanism." IEEE International Conference on Communications (ICC), (2015).
18. Raghu K. Ganti, Fan Ye, Hui Lei.: "Mobile crowdsensing: current state and future challenges." IEEE Communications Magazine, (2011).
19. Y. Zheng, F. Liu, and H. Hsieh.: "U-Air: When urban air quality inference meets big data." In Proceedings of the 19th ACM SIGKDD international conference on Knowledge discovery and data mining (KDD), pp. 1436-1444 (2013).
20. V. Coric and M. Gruteser.: "Crowdsensing Maps of On-street Parking Spaces." In Proceedings of the 2013 IEEE International Conference on Distributed Computing in Sensor Systems (DCOSS), pp. 115-122 (2013).
21. B. Guo, H. Chen, Z. Yu, X. Xie, S. Huangfu .: "A Mobile Crowdsensing System for Cross-Space Public Information Reposting, Tagging, and Sharing." IEEE Transactions on Mobile Computing, (2015).
22. C. Chen, Y. Wang .: "SPARC: Strategy-Proof Double Auction for Mobile Participatory Sensing." Cloud Computing and Big Data (CloudCom-Asia) , (2013).
23. Ling Tang, Shibo He, Qianmu Li .: "Double-Sided Bidding Mechanism for Resource Sharing in Mobile Cloud." IEEE Transactions on Vehicular Technology, pp. 1798-1809 (2017)
24. He Huang, Yu Xin, Yu-E Sun .: "A Truthful Double Auction Mechanism for Crowdsensing Systems with Max-Min Fairness." Wireless Communications and Networking Conference (WCNC) , (2017)